\documentclass{appolb}
\usepackage{graphicx}
\usepackage{amssymb}

\begin{document}
\renewcommand{\runhead}{Intrinsic transverse momentum and parton 
correlations \ldots}
\title{Intrinsic transverse momentum and parton correlations \\
from nonperturbative short-range interactions\thanks{Proceedings 
of Light Cone 2012, Cracow, 
8--13 July 2012}}
\author{P.~Schweitzer${}^1$, M.~Strikman${}^2$, C.~Weiss${}^3$ 
\address{${}^1$ Dept.\ of Physics, University of Connecticut,
Storrs, CT 06269, USA \\
${}^2$ Dept.\ of Physics, Pennsylvania State Univ.,
University Park, PA 16802, USA \\[-.3ex]
${}^3$ Theory Center, Jefferson Lab, Newport News, VA 23606, USA}}
\maketitle
\begin{abstract}
We summarize recent progress in understanding the effects of nonperturbative 
short--range interactions in QCD on the nucleon's partonic structure at 
a low scale: (a) Sea quarks have intrinsic transverse momenta up
to the chiral symmetry--breaking scale $\rho^{-1} \sim 0.6\, \textrm{GeV}$,
much larger than those of valence quarks. (b) Sea quarks in the nucleon 
light-cone wave function exist partly in correlated in pairs of transverse 
size $\rho$ with sigma and pi--like quantum numbers and a distinctive
spin structure ($L = 1$ components). The effects are demonstrated in an 
effective model of the low--energy dynamics resulting from chiral symmetry 
breaking in QCD. They have numerous implications for the $P_T$ 
distribution of hadrons in semi-inclusive DIS and multiparton processes 
in high--energy $pp$ collisions.
\end{abstract}
%
%
Describing the transition between the short--distance regime of 
asymptotic freedom and long--distance hadronic structure is perhaps
the main challenge in practical applications of QCD. A basic observation 
is that non--perturbative effects become important already at distances 
much smaller than the typical hadronic size $R \sim 1\, \textrm{fm}$. 
In the usual approach to QCD based on equal--time quantization and
Euclidean correlation functions these effects lead to a non--trivial 
ground state and are often referred to as ``vacuum structure;'' 
however, their significance really lies in the 
existence of short--range nonperturbative interactions which can 
manifest themselves in hadronic structure in multiple ways.
Specifically, dynamical chiral symmetry breaking
in QCD is caused by non--perturbative interactions over a range 
$\rho \sim 0.3\, \textrm{fm}$ (see Fig.~\ref{fig:scales}), defined 
by the typical size of the topological gauge field fluctuations 
creating the chiral condensate. The existence of this 
non--perturbative short--distance scale has far--reaching consequences 
for hadron structure. It is the foundation of the ``constituent quark'' 
picture explaining many aspects of static nucleon properties and 
low--energy interactions. An outstanding question 
is what the chiral symmetry--breaking interactions at the scale $\rho$ 
imply for the nucleon's partonic structure, i.e., the light--cone 
momentum distributions of quarks, antiquarks and gluons measured in 
deep--inelastic scattering (DIS) and other high momentum--transfer 
processes. A recent study \cite{Schweitzer:2012hh} found two 
striking manifestations:
%
%
\begin{figure}
\centerline{\includegraphics[width=8.5cm]{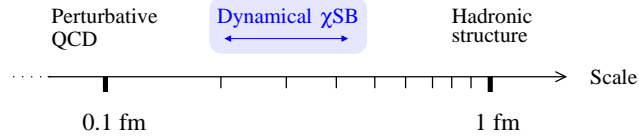}}
\caption{Scale of dynamical chiral symmetry breaking in QCD.}
\label{fig:scales}
\end{figure}
\begin{itemize}
\item[(a)] Sea (or non--valence) quarks in the nucleon at a low scale have 
trans\-verse momenta extending up to the inverse chiral symmetry--breaking 
scale, $p_T \sim \rho^{-1}$, much larger than those of valence quarks,
which are of the order of the inverse hadronic size $R^{-1}$. 
\item[(b)] Sea quarks in the nucleon's light--cone wave 
function are partly correlated in pairs of transverse size $\rho \ll R$
with sigma and pi--like quantum numbers and a distinctive spin structure
(including $L = 1$ components), amounting to non--perturbative 
short--range correlations of partons at a low scale. 
\end{itemize}
These effects were demonstrated in a dynamical model using chiral
constituent quarks as effective degrees of freedom; because they
rely only on qualitative features of the non--perturbative dynamics 
($\rho \ll R$) they are expected to hold also in QCD in a properly 
defined context. They have potential implications for the understanding 
of transverse nucleon structure (role of chiral symmetry breaking, 
valence vs. sea quarks), the effectiveness of QCD factorization 
in DIS with identified transverse momenta 
(natural scales, QCD evolution), and the phenomenology of scattering
processes sensitive to intrinsic transverse momentum and parton
correlations (semi--inclusive DIS, multiparton processes in $pp$
collisions). In this note we summarize the main points; for details 
we refer to the original article \cite{Schweitzer:2012hh}.

%
%
\begin{figure}
\begin{tabular}{ll}
\hspace{.1cm}
\parbox[c]{4.8cm}{\includegraphics[width=4.5cm]{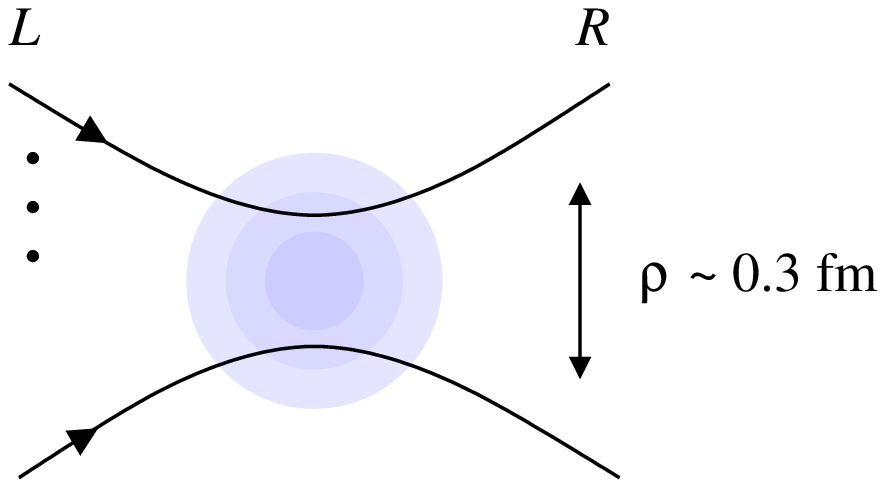}}
&
\parbox[c]{6.3cm}{\includegraphics[width=6.3cm]{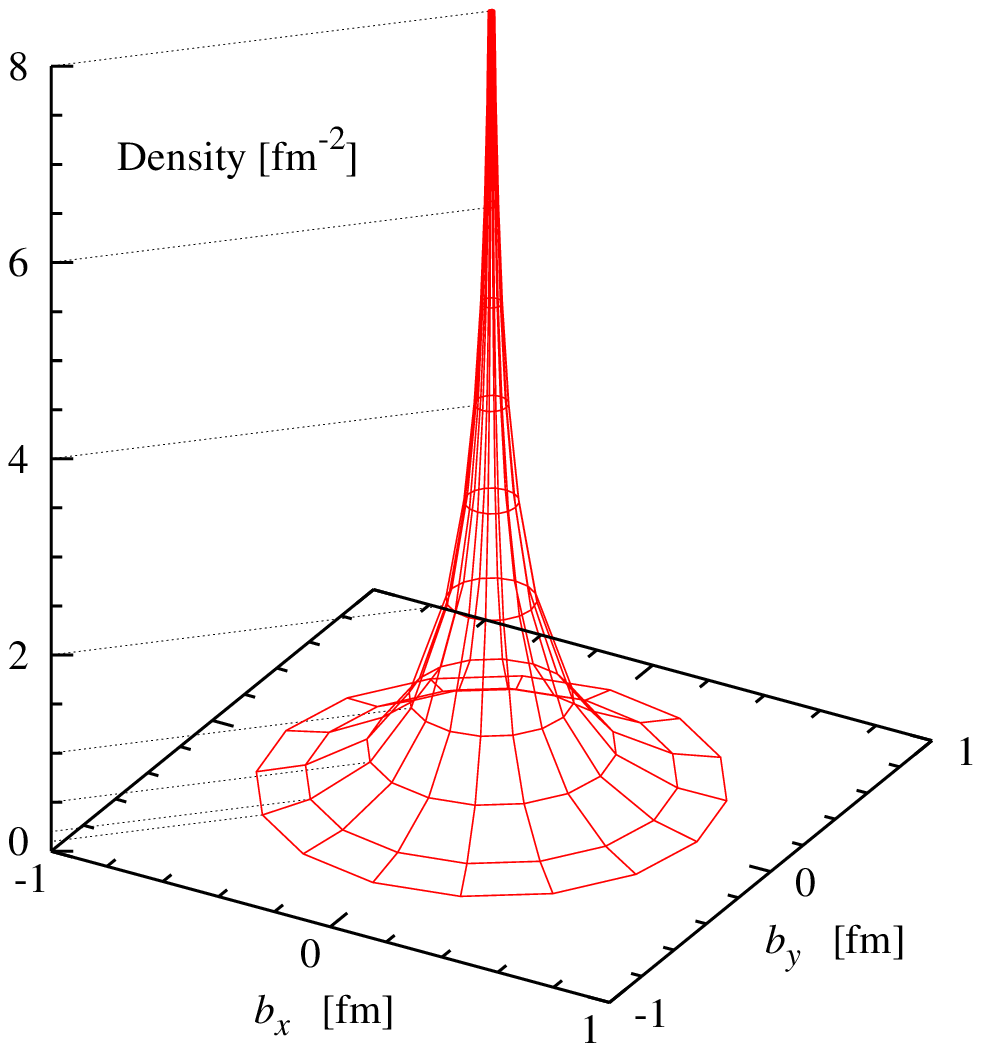}}
\\[-3ex]
{\scriptsize (a)} & {\scriptsize (b)}
\\[1ex]
\end{tabular}
\caption{(a) Chiral symmetry breaking by topological gauge
fields in Euclidean QCD. (b) Transverse charge density in the 
pion obtained from $e^+e^-$ data \cite{Miller:2010tz}. The large density at 
$b \ll 1\, \textrm{fm}$ is due to small--size configurations
likely generated by short--range nonperturbative interactions.}
\label{fig:chiralflip}
\end{figure}
\textit{Chiral symmetry breaking from short--range interactions.}
The short--range character of the nonperturbative interactions causing
the dynamical breaking of chiral symmetry is seen in numerous
theoretical and phenomenological observations. The most direct theoretical
evidence comes from studies of the size distribution of chirality--flipping 
topological gauge fields in lattice simulations of Euclidean QCD (see 
Fig.~\ref{fig:chiralflip}a) \cite{Diakonov:2002fq}.
The instanton vacuum, an approximate realization of this 
mechanism, uses an average size 
$\rho \sim 0.3\, \textrm{fm}$ \cite{Diakonov:2002fq}. 
Other evidence comes from the large value of the 
``average quark virtuality'' in the chiral condensate, 
$m_0^2 = 2 \langle\bar\psi \nabla^2 \psi\rangle/\langle\bar\psi \psi\rangle
\gtrsim 1\, \textrm{GeV}^2$, obtained in lattice QCD 
and the instanton vacuum \cite{Polyakov:1996kh}. Phenomenologically,
the short range of chiral symmetry--breaking interactions is seen in 
the success of the constituent quark picture of hadron 
structure \cite{Diakonov:1995zi}. The ``superconducting'' quark 
model \cite{Ebert:1982pk} and models based on Dyson--Schwinger equations 
of QCD \cite{Roberts:2012sv} describe chiral symmetry breaking as the 
generation of a dynamical quark mass of 0.3--0.4 GeV from effective 
interactions with a range $\ll 1\, \textrm{fm}$ (this aspect 
is not often emphasized).

\textit{Pion structure as example.}
Besides generating a dynamical quark mass, the short--range interactions
associated with chiral symmetry--breaking induce dynamical correlations 
which manifest themselves when probing had\-ron structure at 
distance scales $\sim \rho$. Particularly interesting is the structure 
of the pion, the Goldstone boson of chiral symmetry breaking. A recent 
dispersion analysis using timelike form factor data from $e^+e^-$ 
annihilation experiments found a large transverse charge density 
in the pion at distances $b \ll 1\, \textrm{fm}$ 
(see Fig.~\ref{fig:chiralflip}b) \cite{Miller:2010tz}. Interpreted in 
the context of light--cone wave functions it directly attests to the 
presence of small--size configurations in the pion, as would be produced 
by chiral symmetry--breaking interactions with a range $\rho \ll R$; 
it cannot be explained by end--point configurations ($x \rightarrow 1$) 
at any reasonable normalization point.

%
%
\begin{figure}
\centerline{\includegraphics[width=11cm]{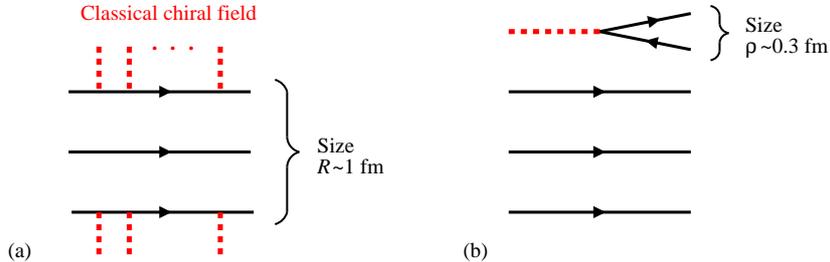}}
\caption{Chiral quark--soliton model of nucleon. The classical chiral 
field (a) binds the valence quarks; (b) creates quark--antiquark pairs.}
\label{fig:chqsm_fields}
\end{figure}
\textit{Chiral quark--soliton model of nucleon.}
The effect of the nonperturbative short--range interactions on the 
nucleon's partonic structure at a low scale can be studied in a schematic
model of the effective dynamics resulting from chiral symmetry--breaking
\cite{Schweitzer:2012hh}. It uses ``constituent'' quarks with a 
dynamically generated mass $M \sim$ 0.3--0.4 GeV as effective degrees 
of freedom below the chiral symmetry--breaking scale. 
The quark mass is accompanied by a coupling to a Goldstone boson
(or chiral) field with strength $M/f_\pi \sim$ 3--4, resulting 
in a strongly coupled field theory that is solved non--perturbatively
in a $1/N_c$ expansion.
The nucleon in this model develops a classical chiral field with
a radius of the order $\sim M^{-1}$, which acts in two ways: it binds the 
valence quarks and creates quark--antiquark pairs out of the vacuum,
which can interact further with the chiral field 
(chiral quark--soliton model, see 
Fig.~\ref{fig:chqsm_fields}) \cite{Diakonov:1987ty}. 
The resulting description is fully 
relativistic and can be studied either in the rest frame, where the 
classical chiral field and the quark orbitals are spherically 
symmetric (``hedgehog''), or in the infinite--momentum frame,
where the fields can be projected on partonic quanta. The light--cone 
wave function of the nucleon in this model is built up from two types 
of configurations:
\begin{itemize}
\item[(a)] Valence quarks in configurations of transverse size 
$R \sim M^{-1}$;
\item[(b)] Quark--antiquark pairs in configurations 
with sizes ranging from the chiral symmetry--breaking scale 
$\rho \ll R$ to the nucleon size $R$. 
\end{itemize}
The chiral quark--soliton model has been used extensively to study the 
nucleon's parton densities at the scale 
$\mu^2 \sim \rho^{-2}$ \cite{Diakonov:1996sr}. 
It predicts a non--trivial antiquark content at this scale as a result
of dynamical chiral symmetry breaking, in agreement with results of 
fits to DIS and other data. The model describes the light--cone momentum
distributions of constituent quarks and antiquarks --- effective degrees 
of freedom which are to be matched with QCD quarks, antiquarks and
gluons at the chiral symmetry--breaking scale $\mu^2 \sim \rho^{-2} \gg M^2$.
The fits show that at this scale $\sim 30\%$ of the nucleon's light--cone 
momentum is carried by gluons; exactly how these appear out of the 
effective degrees of freedom is the subject of on--going study and 
requires detailed information on the embedding of the effective
model in QCD.

%
%
\begin{figure}
\parbox[c]{8cm}{\includegraphics[width=8cm]{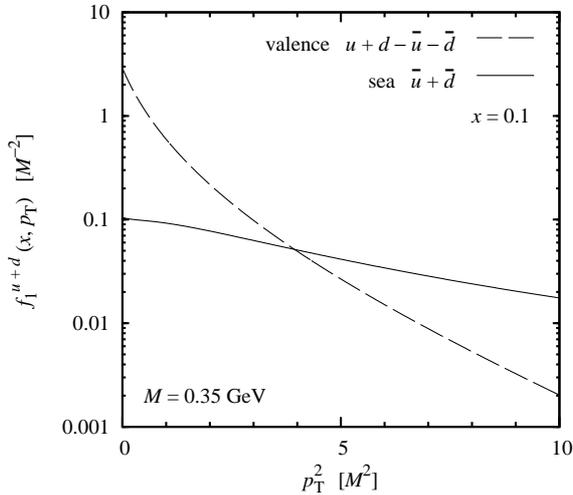}}
\hspace{.2cm}
\parbox[c]{3.7cm}{
\caption{Transverse momentum distribution of valence quarks 
(dashed line) and sea quarks (solid line) in the chiral
quark--soliton model \cite{Schweitzer:2012hh}.}
\label{fig:f1_val_sea}}
\end{figure}
\textit{Transverse momentum distributions.}
The effective model allows us to study also the transverse momentum 
distributions of constituent quarks and antiquarks in the nucleon. 
Fig.~\ref{fig:f1_val_sea} shows the flavor--singlet unpolarized distribution 
$f_1^{u + d}(x, p_T)$ of valence and sea quarks at a representative 
value of $x = 0.1$. The valence quark distribution 
is concentrated at values $p_T^2 \sim \textrm{few}\, M^2$.
The sea quark distribution is qualitatively different and exhibits 
a power--like ``tail'' that extends up to the chiral symmetry--breaking
scale $\rho^{-2}$ (for details see Ref.~\cite{Schweitzer:2012hh}). 
Similar behavior is found in the flavor--nonsinglet polarized 
distribution $g_1^{u - d}(x, p_T)$. 
These features follow from the basic properties of the 
configurations in the light--cone wave function 
(see Fig.~\ref{fig:chqsm_fields}) and represent a direct imprint of 
chiral symmetry breaking on the nucleon's partonic structure. 
They have many potential 
consequences for the modeling of transverse momentum--dependent
(TMD) distributions in QCD and the role of QCD evolution in processes
with identified soft transverse momenta. We note that the 
uncertainties in the matching of the model results with QCD 
presently preclude a fully quantitative interpretation. However, the 
qualitative difference between the valence and sea quark $p_T$ 
distributions follows just from the existence of the two dynamical 
scales $\rho \ll R$ and should hold in QCD in a properly defined context.

\textit{Parton short--range correlations.} Much more insight into the 
role of chiral symmetry breaking can be gained by going beyond the level 
of one--body densities and studying two--particle correlations in
the partonic structure. The effective model predicts short--range 
correlations between constituent quarks and antiquarks in the nucleon's 
light--cone wave function as a direct result of the dynamical mechanism 
by which quark--antiquark pairs are created by the classical field 
(see Fig.~\ref{fig:chqsm_fields}b). The correlated pairs have
either scalar--isoscalar ($\Sigma$) or pseudoscalar--isovector ($\Pi$) 
quantum numbers; the pair's internal light--cone wave function 
involves components with $L = 1$ which dominate at $p_T^2 \gg M^2$
and generate the ``tail'' in the sea quark $p_T$ distribution
(see Fig.~\ref{fig:f1_val_sea}). A very gratifying result is that
at $p_T^2 \gg M^2$ the wave functions of the $\Sigma$-- and $\Pi$--type 
pairs become identical, $|\Psi_\Sigma|^2 = |\Psi_\Pi|^2$, amounting
to ``restoration of chiral symmetry'' at the scale $\rho^{-2}$.
These parton short--range correlations exhibit many similarities
with short--range $NN$ correlations in nuclei \cite{Arrington:2011xs}. 
Note that the fraction of sea quarks in the nucleon with 
momenta $p_T^2 \gg M^2$ is not
small, as can be seen from Fig.~\ref{fig:f1_val_sea}.

\textit{Experimental tests.} The nonperturbative effects in the nucleon's 
partonic structure could potentially be tested in (a)
semi--inclusive DIS discriminating between hadrons produced in scattering 
from quarks and antiquarks; (b) correlations between hadrons in the current 
and target fragmentation regions sensitive to intrinsic $p_T$. Realistic 
projections require detailed modeling of pQCD radiation, final--state 
interactions, and quark fragmentation, and are the object of on--going 
study. Other processes potentially affected by non--perturbative parton 
correlations are exclusive meson production at $W \sim$ few GeV and 
multiparton processes in high--energy $pp$ scattering. 

{\scriptsize Notice: Authored by Jefferson Science Associates, 
LLC under U.S.\ DOE Contract No.~DE-AC05-06OR23177. The U.S.\ Government 
retains a non--exclusive, paid--up, irrevocable, world--wide license to 
publish or reproduce this manuscript for U.S.\ Government purposes.}
\end{document}